\documentclass{article}
\usepackage{frascatiphys}
\begin{document}
\title{THE THEORY OF PENTAQUARKS
}
\author{Harry J. Lipkin\\
{\em Department of Particle Physics, 
Weizmann Institute of Science,} \\ 
{\em Rehovot, Israel}\\ 
and\\
{\em School of Physics and Astronomy} \\
{\em Raymond and Beverly Sackler Faculty of Exact Sciences} \\
{\em Tel Aviv University, Tel Aviv, Israel}\\
and\\
{\em High Energy Physics Division, Argonne National Laboratory}\\
{\em Argonne, IL 60439-4815, USA} 
}
\maketitle
\baselineskip=11.6pt
\begin{abstract}
Is there a theory or good experimental evidence?
 Bj's question: Pentaquark is created by $e^+e^-$. $2q + q \rightarrow$ Baryon 
 $; ~ ~ ~ 2q + \bar q \rightarrow $ Triquark$
; ~ ~ ~ 2q + $Triquark $ \rightarrow $Pentaquark Does it
live long onough to be observable?
Basic physics of constituent quarks and flavor antisymmetry.
Report of $\Theta^+$ violating flavor antisymmetry indicates need
for two-cluster model..
Ball in Experimental Court - 
Some experiments see $\Theta^+$; others don't.
Possible production mechanisms present in some experiments, absent in
others; e.g. via $N^*(2.3 \ \rm{GeV}) \rightarrow \Theta^+ + \bar K$? 
\end{abstract}
\baselineskip=14pt
\section{QCD Guide to the search for exotics}

\subsection{Words of Guidance from  Eugene Wigner's Wisdom}
With a few free parameters I can fit an elephant. 

\noindent With a few more I can make him wiggle his trunk

Wigner's response to questions about a particular theory he did not like was:

\noindent``I think this theory is wrong. But  the old Bohr - Sommerfeld quantum
theory was also wrong..  Could we have reached the
right theory  without it?

\subsection {BJ's question in 1986}
 
In $e^+ e^- $ annihilation a created $q \bar q$ fragments into hadrons.
$q + \bar q \rightarrow$ meson;  $2q + q \rightarrow$ baryon. But 
$2q + \bar q \rightarrow$ Triquark and 
$2q + $Triquark $ \rightarrow$ Pentaquark..

\noindent BJ asked whether quark model says such state.is bound
or lives long enough to be observable as hadron resonance.
Listening to BJ usually pays off.

\subsection{Crucial role of color-magnetic interaction}
\begin{enumerate}

\item QCD motivated models\cite{DGG} show same color-electric interaction  
for large 
multiquark states 
and separated hadrons and no  binding.  Only 
short-range
color-magnetic interaction  produces binding.

\item  Jaffe\cite{Jaffe} extended DGG model\cite{DGG} with one-gluon-exchange color factor to
multiquark sector in a single cluster or bag model, defined $(\bar qq)_8$ and
$(qq)_6$ interactions and explained why lowlying exotics not observed

\item Hyperfine ineraction suggested search for $H$ dibaryon\cite{Jaffe} 
 $uuddss$  and
anticharmed strange pentaquark\cite{Pent97}   $(\bar cuuds)$ (1987)

\end{enumerate}

\subsection{Flavor antisymmetry principle - removes leading exotics}

The Pauli
principle requires flavor-symmetric quark pairs 
to be antisymmetric in color
and spin at short distances. Thus the short-range color-magnetic interaction is always
repulsive between flavor-symmetric pairs. Best candidates for multiquark binding have minimum number of
same-flavor pairs

\begin{enumerate}

\item Nucleon has only one same-flavor pair;  
$\Delta^{++} (uuu)$ has three.   

\item Extra two same-flavor pairs costs 300 Mev . 

\item Deuteron  separates six same-flavor pairs into two nucleons
 
 Only two same-flavor pairs  feel short range repulsion.  

\item $H (uuddss)$ has three same-flavor pairs. Optimum for light 
quark dibaryon

\item The $(uuds \bar c)$ pentaquark has only one same-flavor pair

\item $\Theta^+$ $(uudd \bar s)$ has two same-flavor pairs, more than 
$(uuds \bar c)$.
\end{enumerate}

\noindent Quark model calculations told experimenters "Look for  $\bar c (uuds )$
not $\Theta^+$. 

 Ashery's E791 search for $\bar c uuds$ found events\cite{E791Col}; not  
 convincing enough. 

Better searches for this pentaquark are needed; e.g.
searches with good vertex detectors and good particle ID\cite{Pent97}...

Any proton emitted from secondary vertex is interesting.
One gold-plated event  not a known baryon is enough; No
statistical analysis needed.

\section{ The 1966 basic physics of hadron spectroscopy -  Sakharov-Zeldovich,
Nambu and beyond}

\subsection{Sakharov-Zeldovich (1966)}
 
Sakharov and Zeldovich noted that the $\Lambda$ and $\Sigma$ are made of same
quarks and asked why their  masses are different. Their answer was that a
unified two-body hyperfine interaction not only answers this question but  led
to a unified mass formula for  both meson and baryon  ground states mesons and
baryon masses  and showed that all are made of same quarks\cite{SakhZel}    

\begin{equation}
M = \sum_i m_i + \sum_{i>j}  
{{\vec{\sigma}_i\cdot\vec{\sigma}_j}\over{m_i\cdot
m_j}}\cdot v^{hyp}_{ij}
\label{sakhzel}
\end{equation}

Using (\ref{sakhzel})
Sakharov and Zeldovich noted that 
both the mass difference $m_s-m_u$ between strange and nonstrange quarks and
the flavor dependence of their hyperfine splittings 
(later related\cite{DGG} to the mass ratio $m_s/m_u$) have the same values 
when calculated from baryon masses and meson masses\cite{SakhZel},
along with the comment that the masses
are of course effective masses\cite{Postcard}:
\begin{eqnarray}
& \langle m_s-m_u \rangle_{Bar}&= 
M_\Lambda-M_N=177\,{\rm MeV}\nonumber \\[4pt]
&{}\langle m_s-m_u \rangle_{mes}& =
{{3(M_{K^{\scriptstyle *}}-M_\rho )
+M_K-M_\pi}\over 4} =180\,{\rm MeV} \\[4pt]
& \langle m_s-m_u \rangle_{Bar}&=
{{M_N+M_\Delta}\over 6}\cdot
\left({{M_{\Delta}-M_N}\over
{M_{\Sigma^{\scriptstyle *}}-M_\Sigma}} - 1 \right)
=190\,{\rm MeV}
\nonumber \\[4pt]
&{}\langle m_s-m_u \rangle_{mes}&=
 {{3 M_\rho + M_\pi}\over 8}
\cdot
\left({{M_\rho - M_\pi}\over{M_{K^*}-M_K}} - 1 \right)
= 178\,{\rm MeV}
\, ,
\label{ight:mass}
\end{eqnarray}

 The same value $\pm 3\%$ for $m_s-m_u$ is obtained from 
four independent calculations. The 
same approach for $m_b-m_c$ gives
\begin{eqnarray}
&& \langle m_b-m_c \rangle_{Bar}= M(\Lambda_b)-M(\Lambda_c) =3341 \,{\rm MeV}
\nonumber \\[4pt]
&& \langle m_b-m_c \rangle_{mes} =
{{3(M_{B^{\scriptstyle *}}-M_{D^{\scriptstyle *}})
+M_B-M_D}\over 4} =3339 \,{\rm MeV} 
\label{heavy:m}
\end{eqnarray}

The same value $\pm 2.5\%$ for the
ratio ${{m_s}\over{m_u}}$ is obtained from  
meson and baryon masses.
\begin{equation}
 \left({{m_s}\over{m_u}}\right) =
{{M_\Delta - M_N}\over{M_{\Sigma^*} - M_\Sigma}} = 1.53 =
{{M_\rho - M_\pi}\over{M_{K^*}-M_K}}= 1.61
\end{equation}
  
DeRujula, Georgi and Glashow\cite{DGG} in 1975
used
QCD arguments to relate hyperfine splittings to quark masses and baryon
magnetic moments. This led to remarkable agreement with experiment including
three magnetic moment predictions with no free parameters
\begin{eqnarray}
&&{} \mu_\Lambda=
-0.61
\,{\rm n.m.}=
-{\mu_p\over 3}\cdot {{m_u}\over{m_s}} =
-{\mu_p\over 3} {{M_{\Sigma^*} - M_\Sigma} \over{M_\Delta - M_N}}
=-0.61 \,{\rm n.m.}
\nonumber \\[4pt]
&&{}
\mu_p+\mu_n= 0.88 \,{\rm n.m.}
={M_{\scriptstyle p}\over 3m_u}
={2M_{\scriptstyle p}\over M_N+M_\Delta}=0.865 \,{\rm n.m.}
\nonumber \\[4pt]
&&{}
-1.46 =
{\mu_p \over \mu_n} =
-{3 \over 2}\, ,
\label{mag:mom}
\end{eqnarray}

\subsection{Two Hadron Spectrum 
puzzles -Why $qqq$ and $q\bar q$ ? }

\begin{enumerate}
\item The Meson-Baryon Puzzle - The $qq$ and $\bar qq$ forces must be 
peculiarly related
to bind both mesons and baryons. It cannot be a vector interaction giving 
equal and opposite
forces, nor a scalar or tensor giving equal attractions for both.

\item Exotics Puzzle - No low-lying hadrons with exotic quantum numbers have 
been observed; e.g. no $\pi^+ \pi^+$ or $K^+ N$ bound states.
\end{enumerate} 

Nambu solved both puzzles\cite{Nambu} in 1966 by introducing the color degree
of freedom and  a two-body interaction from a non-abelian gauge theory  with
the color-factor of one-gluon exchange. This both related mesons and baryons
and eliminated exotics.

A unified treatment of $qq$ and $\bar qq$ interactions binds both mesons and
baryons with the same forces.  
Only $qqq$ and $q\bar q$ are stable in any single-cluster model with color
space factorization. Any color singlet cluster that can break up into two color singlet 
clusters loses no color electric energy and gains kinetic energy.
The Nambu color factor does not imply dynamics of one-gluon exchange. 
Higher order diagrams can have same color factor

Looking beyond bag or single-cluster models for possible molecular bound
states  Lipkin(1972) showed that the color-electric potential energy  could be
lowered in potential models by introducing color-space correlations; e,g,
$q\bar qq \bar q$ at  corners of a square, but not enough to compensate for the
kinetic energy\cite{LipkTriEx}

\subsection{Important systematics in the experimental spectrum}
A large spin-dependent interaction $\approx$ 300 MeV
but a very weak interaction $\approx$ 2 MeV binding normal hadrons.
\begin{equation}
 M(\Delta) - M(N) \approx 300 MeV \gg M(n) + M(p) - M(d)
 \approx 2 MeV 
\end{equation}
\subsection{Conclusions from basics}

The low-lying hadron spectrum is described by a linear effective mass term and 
a hyperfine
interaction with a one-gluon exchange color factor.

The $(\bar q q)$ and $(qqq)$ states behave like neutral atoms
with a strong color electric field inside hadrons and none outside. 
No molecular bound states arise in the simplest cases.
A strong spin-dependent interaction is crucial to understanding the spectrum

Only color singlet and $3^*$ color factors
arise in $(\bar q q)$ and $(qqq)$. The low-lying hadron spectrum provides  no
direct experimental information on $(\bar q q)_8$ and $(qq)_6$  interactions
needed for multiquark exotic configurations.

\subsection{What can QED teach us about QCD?}

QCD is a Great Theory, but nobody knows how to connect it with experiment or
which approximations are good. We need  to construct instructive simplified
models. I often recall the response by Yoshio Yamaguchi at a seminar at the
Weizmann Institute in 1960 when asked if there had been any thought at CERN
about a possible breakdown of QED at small distances: ``No. .   Many
calculations. No thought."

What can we learn from QED; a Great Theory
that everyone knows how to connect with experiment?
We know how isolated free electrons behave and carry currents. 
But nobody could explain the fractional Hall effect.until Robert Laughlin told
us  the Hall Current is not carried by single electrons! 
It is carried by quasiparticles related to electrons by a complicated
transformation.

Nobody has ever seen an isolated free quark. 
Current quark fields appear in the Standard Model Lagrangian.
But experiments tell us that baryons are $qqq$ and mesons are $q \bar q$  
and these are not the quarks that appear in the QCD Lagrangian.

Nobody knows what these quarks are.
Are they complicated quasiparticles related to current quarks
by a complicated transformation?.
Is Hadron Spectroscopy Waiting for Laughlin?
Does QCD need another Laughlin to tell us what constituent quarks are?

\section {The $\Theta^+$ was reported! A Two-cluster Model?} 

\subsection{Following Wigner's Guidance  to Understand QCD and the
Pentaquark}

One good wrong model that stays away from 
 free parameters and may  teach us something: 
a two-cluster $P$-wave $(ud)$  diquark-$(ud \bar s)$ triquark  
model\cite{OlPenta,NewPenta} for the $\Theta^+$ 
that separates $uu$ and $dd$ pairs and eliminates 
their short range repulsive interaction...
Its hidden-strangeness $N^*$ partner keeps the same triquark with the  
$(us)$ and $(ds)$ $SU(3)$ partners of 
the $(ud)$ diquark. Its mass is roughly\cite{varxist}
\begin{equation}
M[N^*(1775)] \approx M(\Theta^+) + M(\Lambda) - M(N) + 
{3\over 4}\cdot [M(\Sigma)-M(\Lambda)] \approx 1775 \,{\rm MeV}   
\end{equation}
 
 \subsection {The skyrmion model}
Experimental search motivated by another wrong model. 
Skyrmion model has no simple connection with quarks 
except by another wrong model.  The $1/N_c$ expansion
invented\cite{LipHouch} pre-QCD to explain absence of free quarks.

The 
binding Energy of $q \bar q$ pairs into mesons $ E_M \approx g^2 N_c$.

At large $N_c$ the cross section for meson-meson scattering breaking up a meson
into its constituent quarks is
\begin{equation}
\sigma [MM \rightarrow M + q + \bar q] \approx g^2  {{E_M}\over{N_c}}
\approx 0 
\end{equation}
But ${1\over {N_c}} ={1\over 3}$;  ${\pi\over {N_c}} \approx 1$  
This is NOT A SMALL PARAMETER!

\section {Experimental contradictions about the $\Theta^+$}  

Some experiments\cite{Nakano:2003qx,
Kubarovsky:2003fi,NakanoNSTAR2004,Troyan:2004wp}
see the $\Theta^+$; others\cite{BES, DELPHI} definitely do not.
Further analysis is needed to check presence of specific production mechanisms
in experiments that see it and their absence in those that
do not\cite{cryptopen}. No theoretical model addresses this question.
Comprehensive
review\cite{jenmalt} analyzes different models..  

\subsection{Production via decay of a
cryptoexotic $N^*(2400)$}

The 
reported\cite{Kubarovsky:2003fi,BurkertNSTAR2004,Landsberg:1999wn,NA49Theta}
$N^*(2400)$ can be the $D$-wave excitation of the $N^*(1775)$.with a 
$(ds)$ diquark in a $D$-wave
with the 
 same $ud\bar s$ triquark. Its dominant decays would be 
$N^*(2400) \rightarrow K^- \Theta^+$  via the diquark transition 
$ds \rightarrow ud + K^-$.and 
$N^*(2400) \rightarrow \pi^- N^*(1775)^+ \rightarrow \pi^- \Lambda K^+$  via 
$ds \rightarrow us + \pi^-$. 

Decays like
$\Lambda K$and $\Sigma K$ would be suppressed by the centrifugal barrier
forbidding a quark in the triquark from joining the diquark.

Some experimental checks of this mechanism are:

\begin{enumerate}
 \item
Experiments which see the $\Theta^+$ and have sufficient energy for
producing the $N^*(2400)$ should look for an accompanying $K^-$ or $K_s$ and
examine the mass spectrum of the $K^-\Theta^+$ and $K_s \Theta^+$
systems.
\item 
Experiments should look for 
$N^*(2400) \rightarrow \pi^- N^*(1775)^+ \rightarrow \pi^- \Lambda K^+$  .
\item 
Experiments searching for the $\Theta^+$ should check possible
production of a $K^-\Theta^+$ or $K_s \Theta^+$ resonance in the 2.4
GeV region. $B$-decay modes 
suggested for pentaquark searches\cite{rosner,Browder:2004mp}
 would not produce this 2.4 GeV $N^*$.
Similar considerations should be applied
to searches in $e^+e^-$ and $\gamma\gamma$ like those proposed in
Ref.~\cite{Armstrong:2003zc}.

\item The other $N^*(2400)$ decay modes.; e.g. $K \Lambda$, $K \Sigma$,
$K\Sigma^*$, $\phi N$, are suppressed by the centrifugal barrier in the
D-wave diquark-triquark model but may be appreciable..
Finding them would would give further evidence
for this model for pentaquark production.
The relative branching ratios would also provide information about the
structure of th $N^*(2400)$. 

\end{enumerate}

\subsection{Angular distribution tests for production mechanisms} 

\begin{enumerate}
\item
The angular distribution of the kaon emitted with the $\Theta^+$ in
$\gamma p \rightarrow \bar K^o \Theta^+$ \cite{CLAS-Trento}
carries interesting information.
Production from a cryptoexotic $N^*$, gives no
forward-backward kaon asymmetry. Meson exchange
gives forward peaking. 
Baryon exchange gives backward peaking,
produces the $\Theta^+$ equally by photons on protons and neutrons.
and the same baryon exchange should be seen\cite{Karliner:2004qw}
in \ $\gamma n \rightarrow K^-  \Theta^+$..
\item
The more complicated angular distributions in 
\ $\gamma p \rightarrow \pi^+ K^- K^+ n$
\ \cite{Kubarovsky:2003fi}
may still carry interesting information. 

 All the above
discussion for  $\gamma p \rightarrow \bar K^o
\Theta^+$ applies to the angular distribution of a $\bar K^*$.in
\ $\gamma p \rightarrow \bar K^{*o} \Theta^+ \rightarrow \pi^+ K^-\Theta^+$. 
Models\cite{Karliner:2004qw} with a suppressed $N K \Theta^+$ coupling
 relative to $N K^*
\Theta^+$ predict stronger $\Theta^+$
production with a backward $K^*$   than
with a backward kaon..
In $\gamma p \rightarrow \pi^+ N^*  \rightarrow \pi^+ \Theta^+ K^- $
\cite{Kubarovsky:2003fi}, the pion goes forward and
everything else is in the target fragmentation region.
\cite{BurkertNSTAR2004}..
\end{enumerate}
\subsection{Other experimenal considerations} 

\begin{enumerate}

\item

Search for exotic positive-strangeness baryon exchange in normal
nonexotic reactions. The baryon exchange diagram\cite{CLAS-Trento} for
$\Theta^+$ photoproduction with an outgoing kaon is simply related to backward
$K^-p$ charge-exchange \cite{Karliner:2004qw}. The lower $K N \Theta^+$
vertices are the same; the upper vertex is also $K N \Theta^+$ for $K^-p$
charge-exchange but $\gamma \Theta^+ \Theta^+$ for $\Theta^+$
photoproduction.
If this diagram contributes appreciably to $\Theta^+$ photoproduction, 
the contribution of the  $K N \Theta^+$ vertex is appreciable
and should also contribute appreciably to backward $K^- p$ charge-exchange.
Some previously ignored backward $K^- p$ charge-exchange data may be available.

\item The baryon and $\bar s$ constituents of the $\Theta^+$   are
already initially present in low-energy photoproduction experiments in the
target baryon and the $\bar s$ component of the photon. In experiments where
baryon number and strangeness must be created from gluons, the cost of baryon
antibaryon and $s\bar s$ production by gluons must be used to
normalize the production cross section in comparison with the photoproduction
cross sections; e.g. from  baryon-antibaryon production and $s \bar s$
production data in the same experiment that does not see the $\Theta^+$.

\end{enumerate}

\section{Acknowledgements}
The original work reported in this talk was in collaboration with Marek
Karliner.
This work was partially
supported by the U.S. Department
of Energy, Division of High Energy Physics, Contract W-31-109-ENG-38

\end{document}